\documentclass[letterpaper,english,reprint,prb]{revtex4-1}
\usepackage[T1]{fontenc}
\usepackage[latin9]{inputenc}
\usepackage{babel}
\usepackage{array}
\usepackage{textcomp}
\usepackage{amsmath}
\usepackage{amssymb}
\usepackage{graphicx}
\usepackage{epstopdf-base}
\usepackage{multirow}
\usepackage{rotating}
\usepackage{graphics}
\usepackage{xcolor}
\usepackage[unicode=true,pdfusetitle,bookmarks=true,bookmarksnumbered=false,bookmarksopen=false, breaklinks=true,pdfborder={0 0 1},backref=false,colorlinks=false] {hyperref}

\makeatletter

\pdfpageheight\paperheight
\pdfpagewidth\paperwidth

\providecommand{\tabularnewline}{\\}

\usepackage{babel}

\pdfpageheight\paperheight
\pdfpagewidth\paperwidth

\providecommand{\tabularnewline}{\\}

\makeatother

\begin{document}

\preprint{APS/123-QED}

\title{van der Waals coefficients of the multi-layered MoS$_2$  with alkali metals}

\author{Shankar Dutt}
\author{Sukhjit Singh}
\author{A. Mahajan}
\author{Bindiya Arora}
\email{bindiya.phy@gndu.ac.in}

\affiliation{Department of Physics, Guru Nanak Dev University, Amritsar, Punjab-143005, India}

\author{B. K. Sahoo}

\affiliation{Atomic, Molecular and Optical Physics Division, Physical Research Laboratory, Navrangpura, Ahmedabad-380009, India}


\begin{abstract}
The van der Waals coefficients and the separation dependent retardation functions of the interactions between the atomically thin films of the
multi-layered transition metal molybdenum disulfide (MoS$_2$) dichalcogenides with the alkali atoms are investigated. First, we determine the 
 frequency-dependent dielectric permittivity and intrinsic carrier density values for different layers of MoS$_2$  by
adopting various fitting models to the recently measured optical data reported by Yu and co-workers [Sci. Rep. {\bf 5}, 16996 (2015)] using 
spectroscopy ellipsometry. Then, dynamic electric dipole polarizabilities of the alkali atoms are evaluated very accurately by employing the 
relativistic coupled-cluster theory. We also demonstrate the explicit change in the above coefficients for different number of layers. These
studies are highly useful for the optoelectronics, sensing and storage applications using layered MoS$_2$.
\end{abstract}

\maketitle

\section{\label{sec:level1}Introduction}

Recent advances in the fabrication and synthesis of ultra-thin layered materials with unit cell thickness has boosted the art of continuously 
tailoring the properties of materials. Among these, graphene, a two dimensional (2D) material of carbon atoms, exhibit unique electronic, 
physical and chemical properties. However, zero band gap of graphene restricts the direct application of graphene in the electronic devices. 
This has prompted to search for composite graphene-like materials with a finite band gap. Transition metal dichalcogenides (TMDs) possessing 
the identical lamellar structure of graphite manifest remarkable applications in nano-electronics, sensors, catalytic, energy conversion and 
storage devices. The nature of transition elements present in these materials affects their structures. TMDs containing transition elements 
from the IV-VII groups of the periodic table exhibit layered structures, while those containing transition elements belonging to the VIII-X 
groups show non-layered structures~\cite{han2015}. 

In addition to the homo-layer configurations in the 2D TMDs, the nanoscale heterostructures of TMDs have also been found to be suitable for the 
implementation of novel photonic and electronic devices~\cite{gong2013,komsa2013}. Theoretical studies demonstrate that several 2D TMDs offer 
a plethora of opportunities using lateral and vertical heterostructures due to their tunable broad-range optical bandgap and strong light-matter 
interactions \cite{pksahoo1,pksahoo2}. These heterostructures can be classified into three types on the basis of their band alignments; i.e. symmetric 
(type I), staggered (type II) and broken (type III)~\cite{ozcelik}. All these materials find many applications in high performance devices such 
as light-emitting diodes, photodetectors and transistors~\cite{withers2015,xu2017,nakamura1995,yu2013,zhang2011,sarkar2015,lin2015,koswata2010}. 
Moreover, strong Coulomb interactions and anisotropic dielectric environment lead to the formation of strongly bound excitons, trions, and 
biexcitons in these materials~\cite{you2015, he2014, mak2012}. The hetero-bilayers of TMDs with inter-layer excitons are emerging as novel class 
of long-lived dipolar composite bosons for optoelectronic applications~\cite{kremser2020}. Recently, it has been found that magnetic fields can 
promote the formation of biexciton to create favorable conditions for the formation of multiple exciton complexes, exciton super-fluidity, and 
biexciton condensates to materialize their practical applications~\cite{stevens2018}.

Among TMDs, molybdenum disulfide (MoS$_2$) has emerged as one of the promising next-generation 2D materials with exceptional photonic, non-linear 
and electronic properties, in contrast to its bulk counterpart \cite{Q.H.Wang2012,Chhowalla2013,Li2015} and attracted applications among 
flexible gas sensing \cite{Donarelli2018,Akbari2018} and optoelectronic devices \cite{Mak2010,Brasch2016,Tsai2014,Yin2012,Guo2016,Tongay2012,
Gutuerrez2013,Sundaram2013,Zhao2013,Krishnan2019,Li2018}. In order to design and simulate the next-generation nanoelectronic devices built with 
MoS$_2$, it is important to gain accurate knowledge of its electrical permittivity ($\epsilon$), which is a fundamental property that 
characterizes refractive index, absorption, conductivity, capacitance, and many other intrinsic phenomena of a material \cite{elton}. Owing to 
complexity involved in the determination of $\epsilon$ values, a number of studies on these quantities for different layers of MoS$_2$ have been 
carried out. These investigations report a wide range of values with substantial differences in magnitudes from each other. This can be evident
from the following: Liang et al. \cite{Liang_1971} and Beal et al.~\cite{Beal_1979} presented a reflectivity spectrum of MoS$_2$ and calculated 
$\epsilon$ values by adopting Kramers-Kronig procedures. Liu et al.~\cite{tao} demonstrated that $\epsilon$ can be deduced from the 
absorption spectra. They first extracted out the imaginary part of $\epsilon$ from the absorption spectra, then estimated the real part 
using the Kramers-Kronig relation. Li et al.~\cite{lisl} inferred $\epsilon$ from differential reflection spectra using an effective 
reflection coefficient method. Castellanos-Gomez et al.~\cite{gomez} studied the refractive index of thin MoS$_2$ crystal with the Fresnel 
law and further predicted $\epsilon$ values. Recently, Yu et al.~\cite{yu} have measured $\epsilon$ values as functions of the number of 
layers for a discrete wavelength spectra in the visible region (345 nm to 1000 nm) using spectroscopic ellipsometry technique. Their employed  
method is specially designed to measure optical data very accurately, so it is expected that measured values of $\epsilon$ by Yu et al. are 
more reliable than the above-estimated values using various methods. 

Apart from the electronic properties, knowledge of single atom adsorption with the atomically thin layered surfaces is of great importance for 
many practical applications. For instance, the alkali metal atom adsorption on graphene generally leads to an increase in its Fermi level, that 
has  excellent potential for the field emission applications~\cite{32}. Moreover, Li ion storage capacity of single boron-doped graphene is found 
to be dramatically improved~\cite{33}. It is also known that if alkali atoms are absorbed on a metal surface, the electron and ion emission 
properties of the surface are drastically altered to provide improved applications in thermionics and physical electronics. The intercalations of 
alkali-metal ions (such as Li$^+$, Na$^+$, K$^+$) in 2D-layered MoS$_2$ can induce structural phase changes along with introducing changes in 
their electronic and optical properties~\cite{1aman,2aman,3aman,4aman}. The 2D MoS$_2$ nanoflakes on intercalation with Li$^+$ ions exhibit 
plasmon resonances near-UV and visible regions. These materials have potential applications in the optoelectronics as well as in the plasmonic 
biosensing~\cite{5aman,6aman}. Additional efforts have also been made to manipulate the electronic properties of MoS$_2$ through single-atom 
adsorption~\cite{lifang,Xiao2012,Ataca2011}. The van der Waals (vdW) interactions between atoms and material surfaces are critical for the study
of physical adsorption. The interactions of atoms having lower ionization potentials with MoS$_2$ layers are considered to be crucial
for a large number of possible applications requiring low-energy plasmas and ion beams \cite{Gadzuk1967}. From this point of view, it is
important to fathom vdW interactions among alkali atoms with the material media; especially with the MoS$_2$ layers. 

Motivated by the above developments, we report the vdW interactions between different alkali atoms and MoS$_{2}$ based TMDs. The electrical 
permittivity data required for such calculations have been taken from the ellipsometry measurements of Yu et. al.~\cite{yu}. Calculations of 
dynamic dipole polarizabilities at imaginary frequencies for the respective atoms required for this study are carried out using relativistic 
coupled-cluster (RCC) theory. We qualitatively evaluate the intrinsic carrier density ($N$) for MoS$_2$ layers by fitting the experimental 
permittivity results with Drude-Lorentz (DL) oscillator. The DL model permits extraction of $N$ from MoS$_2$ based TMDs for different 
number of layers. This allows us to examine the effect of $N$ on the interaction coefficients as functions of the number of layers in MoS$_2$. 
We find that the interactions between neutral atoms and MoS$_{2}$ are directly proportional to $N$, and they are maximum for monolayer. They 
decrease up to the 6th layer, thereafter they start increasing in the MoS$_2$ based TMDs. 

The paper is organized as follows. In Sec. \ref{theory}, we present theoretical formulae used to calculate the vdW coefficients and the 
retardation functions, which can be used to describe the nature of the vdW interactions over a wide range of radial distance. In Sec. \ref{pol}, 
we discuss methods used for accurate evaluation of the atomic dynamic dipole polarizabilities. A well suited permittivity model to extract out 
the values of $N$ is discussed in Sec. \ref{model}. It follows by presenting results and discussion in Sec. \ref{results}, before concluding in 
the last section. 

\section{Theory}~\label{theory}

A consistent theory, accounting for the electrical, mechanical and optical properties of materials, to study the vdW interactions 
among various atomic systems and real bodies made of different materials has been given by E. M. Lifshitz and collaborators 
\citep{lifshitzbook,lifshitz1}. The atom-wall interactions can be computed by considering a polarizable particle 
interacting with a surface or a wall as a continuous medium having a frequency-dependent permittivity with real ($\epsilon_r(\omega)$)
and imaginary ($\epsilon_i(\omega)$) parts. In this theory, the interaction potential of vdW interactions between an atom and 
a layered structure or a material plate can be efficiently described by the following formula \citep{lach1,lifshitzbook,Arora2014,
lifshitz1} 
\begin{eqnarray}
U(z)&=& -\frac{\alpha_{fs}^{3}}{2\pi}\intop_{0}^{\infty}d\omega\omega^{3}\alpha_n(\iota\omega) \nonumber \\
 && \times \intop_{1}^{\infty}d\xi e^{-2\alpha_{fs}\xi\omega z}H(\xi,\epsilon_r(\iota\omega)),\label{eq:1}
\end{eqnarray}
where  $\alpha_{fs}$ is the fine structure constant, $z$ is the distance between the atom and the wall, and $\alpha_n(\iota\omega)$ is the 
ground-state dynamic  dipole polarizability of the atom with imaginary argument. The quantity $H(\xi,\epsilon_r(\iota\omega))$, a function of 
Matsubara frequencies $\xi$ and dielectric permittivity $\epsilon_r(\iota\omega)$ of the material wall, is given by
\begin{equation}
H(\xi,\epsilon_r)=\left [ \left ( 1-2\xi^{2} \right )\frac{\xi'-\epsilon_r\xi}{\xi'+\epsilon_r\xi} \right ] +\frac{\xi'-\xi}{\xi'+\xi} \label{eq:2}
\end{equation}
with $\xi'=\sqrt{\xi^{2}+\epsilon_r-1}$. The procedure for the evaluation of  $\epsilon_r(\omega)$ is explained in Refs. 
\citep{Arora2014,lach1,arora-sahoo1}. In our study, the real ($n(\omega)$) and the imaginary ($\kappa(\omega)$) parts of the refractive 
index of MoS\textsubscript{2} are used to evaluate the imaginary parts of the dielectric permittivity of MoS\textsubscript{2} by the 
relation 
\begin{equation}
\epsilon_i(\omega)=2 \ n(\omega) \ \kappa(\omega).\label{eq:8}
\end{equation}
We use the experimental values of $n(\omega)$ and $\kappa(\omega)$ from Ref.~\cite{yu} to obtain the imaginary part of dielectric permittivity 
values. For conveniently carrying out the calculations and 
to predict the number of intrinsic carrier density $N$ (electrons per unit volume) in the MoS$_2$ layers, we determine $\epsilon_i(\omega)$ 
using the DL oscillator model. This procedure has been discussed latter in detail.Further, we evaluate the real values of the dielectric permittivity at the imaginary frequencies by using the Kramers-Kronig 
formula \citep{Kramers} 
\begin{equation}
\epsilon_r(\iota\omega)=1+\frac{2}{\pi}\int_{0}^{\infty}d\omega' \ \frac{\omega'\epsilon_i(\omega')}{\omega^{2}+\omega'^{2}}.\label{eq:9}
\end{equation}
These values are calculated for the MoS$_2$ layers with layer number ranging from 1 to 10. 

The vdW interaction potential can be conveniently expressed by \citep{kharchenko}
\begin{equation}
U(z)=-\frac{C_{3}}{z^{3}}f_{3}(z),\label{eq:3}
\end{equation}
where $f_{3}(z)$ is the retardation function and $C_{3}$ is known as the vdW coefficient, which is defined by 
\begin{equation}
C\text{\ensuremath{_{3}=\frac{1}{4\pi}\int_{0}^{\infty}d\omega\alpha_n(\iota\omega) \vartheta(\iota\omega) }} \label{eq:4}
\end{equation}
with the factor
\begin{equation}
\vartheta(\iota\omega)=\frac{\epsilon_r(\iota\omega)-1}{\epsilon_r(\iota\omega)+1}.\label{eq:5}
\end{equation}
For the perfect conductor $\vartheta\rightarrow 1$, whereas for other materials, $\vartheta$ can be evaluated with the knowledge of their dielectric
permittivities. By adopting a similar approach as in Ref. \citep{Arora2014}, we determine the vdW interaction potential between an atom and a thin 
layer of MoS\textsubscript{2} by using Eq. (\ref{eq:1}), and evaluate the $C_{3}$ coefficient using Eq. (\ref{eq:4}). By combining the $C_{3}$ 
coefficient and the interaction potential, the functional form of $f_{3}(z)$ for the vdW interaction potential is inferred from Eq. (\ref{eq:3}).  

\section{Dynamic polarizabilities of atoms}~\label{pol}

Evaluation of interaction potential $U(z)$ from Eq. (\ref{eq:1}) requires values of  $\alpha_n(\iota\omega)$. The procedure 
for determining accurate values of the dynamic polarizability of an atomic system having a closed core and a valence electron has been already 
described by us in Refs.~\citep{Kaur2015,Arora2012b}. We apply the same procedure here to calculate the dynamic polarizabilities of the ground state
of various alkali atoms considered in this study. In this approach, we divide the total dipole dynamic polarizability in terms of scalar
and tensor components as follows
\begin{equation}
\alpha_n(\iota\omega)=\alpha_n^{(0)}(\iota\omega) + \frac{3M_{J_n}^2-J_n(J_n+1)}{J_n(2J_n-1)}\alpha_n^{(2)}(\iota\omega).
\end{equation}
Here $\alpha_n^{(0)}(\iota\omega)$ and $\alpha_n^{(2)}(\iota\omega)$ are known as the scalar and tensor polarizabilities respectively. They are 
evaluated using the sum-over-states approach as 
\begin{eqnarray}
\alpha_n^{(0)}(\iota\omega)&=& \sum_{k \ne n} W_{n}^{(0)}  \left [\frac{ |\langle \gamma _nJ_n||{\bf D}||\gamma _k J_k \rangle|^2}{E_n -E_k +\iota\omega}\right. \nonumber \\ 
&& \left.+\frac{ |\langle \gamma _n J_n||{\bf D}||\gamma _k J_k \rangle|^2}{E_n-E_k-\iota\omega}\right], \label{scalar}
\end{eqnarray}
and
\begin{eqnarray}
\alpha_n^{(2)}(\iota\omega)&=& \sum_{k \ne n} W_{n,k}^{(2)}  \left [\frac{ |\langle \gamma _n J_n||{\bf D}||\gamma _k J_k \rangle|^2}{E_n -E_k +\iota\omega} \right. \nonumber \\
&& \left.+\frac{ |\langle \gamma _n J_n||{\bf D}||\gamma _k J_k \rangle|^2}{E_n-E_k-\iota\omega}\right] \label{tensor}
\end{eqnarray}
with the coefficients
\begin{eqnarray}
W_{n}^{(0)} &=&-\frac{1}{3(2J_n+1)}, \label{eqp0}  
\end{eqnarray}
and
\begin{eqnarray}
W_{n,k}^{(2)} &=&2\sqrt{\frac{5J_n(2J_n-1)}{6(J_n+1)(2J_n+3)(2J_n+1)}} \nonumber \\ 
& & \times (-1)^{J_n+J_k+1}
                                  \left\{ \begin{array}{ccc}
                                            J_n& 2 & J_n\\
                                            1 & J_k &1 
                                           \end{array}\right\} ,          \label{eqp2}
\end{eqnarray} 
for the electric dipole (E1) reduced matrix elements $\langle \gamma _n J_n||{\bf D}||\gamma _k J_k \rangle$, $J$ denotes the total angular 
momentum, $E$ stands for energy and $\gamma$ represents for the additional quantum numbers of atomic states.

For each component $i=0$ and 2, we divide contributions to polarizability $\alpha_n^{(i)}$ into three parts, based on the correlation contributions 
from different types of electrons, as \citep{arora-sahoo3,Kaur2015,Arora2012b}
\begin{eqnarray}
\alpha_n^{(i)}  &=& \alpha_{n,c}^{(0)} + \alpha_{n,cv}^{(i)} +  \alpha_{n,v}^{(i)}
\label{eq26}
\end{eqnarray}
where $\alpha_{n,c}^{(0)} $, $ \alpha_{n,cv}^{(i)}$ and $ \alpha_{n,v}^{(i)}$ are referred to as the core, core-valence and valence correlation 
contributions, respectively. The $\alpha_{n,c}^{(0)}$ and $\alpha_{n,cv}^{(i)}$ contributions arise from the core-orbitals without considering 
and including interaction with valence orbital, respectively. These contributions are small in the alkali atoms. We, again, divide the $\alpha_{n,v}^{(i)}$ contribution into
two parts; Main -- containing dominant contributions from the low-lying excited states, and Tail -- containing contributions from the remaining
excited states. As seen in the previous studies, major contributions to the polarizabilities of the atomic states in the alkali atoms come from 
$\alpha_{n,v}^{(i)}$ \citep{derevianko1,arora1,derevianko3,arora-sahoo1,arora-sahoo2,Arora2012b,Arora2014a} owing to the dominant contributions 
from the low-lying excited states. Evaluating the Main contribution exclusively has the advantage that uncertainty in its determination can be reduced 
by using excitation energies and reduced E1 matrix elements from the precise measurements wherever available. Contributions from the Tail part are 
estimated approximately using the Dirac-Fock (DF) method. Similarly, the core-valence contribution $\alpha_{n,cv}^{(0)}$ is also obtained using
the DF method, whereas we adopt a relativistic random phase approximation, as discussed in Ref.~\cite{Kaur2015,yashpal}, to evaluate the
$\alpha_{n,c}^{(0)}$ contribution.

For accurate evaluation of the E1 matrix elements involving the ground and low-lying excited states of the considered atoms, we employ the RCC 
theory {\it ansatz}. In this theory, the wave functions of atomic states in an alkali atom can be expressed by \cite{Mukherjee10,bijaya1,bijaya2,ref1,ref2,bijaya3}
\begin{eqnarray}{\label{wav}}
|\Psi_n\rangle &=& e^T\{1+S_n\}|\Phi_n\rangle , \nonumber 
\end{eqnarray}
where $| \Phi_n \rangle = a_n^{\dagger}|\Phi_0\rangle$ with the DF wave function $|\Phi_0\rangle$ of the closed-core of the atom and $a_n^{\dagger}$
denotes the valence orbital in a given state, $T$ is known as the hole-particle excitation operator, which is responsible for exciting electrons from the occupied 
orbitals, and $S_n$ corresponds to the excitation operator involving electron from the valence orbital $n$. In the present work, we have considered 
singles and doubles excitations in the RCC theory (RCCSD method) by expressing 
\begin{equation}
T=T_1+T_2=\sum_{ap}a_p^{\dagger}a_at_a^p+\frac{1}{4}\sum_{abpq}a_p^{\dagger}a_q^{\dagger}a_ba_at_{ab}^{pq}
\end{equation}
and
\begin{equation}
S_n=S_{1n}+S_{2n}=\sum_{n\neq p}a_p^{\dagger}a_ns^p_v+\frac{1}{2}\sum_{bpq}a_p^{\dagger}a_q^{\dagger}a_ba_ns_{nb}^{pq},
\end{equation}
where $t^p_a$ and $t_{ab}^{pq}$ are the amplitudes of the singles and doubles excitations of the $T$ operator, respectively, and $s_n^p$ and 
$s_{nb}^{pq}$ are the amplitudes of the singles and doubles excitations of the $S_n$ operator, respectively. After obtaining atomic wave functions 
in the RCCSD method, we calculate the E1 matrix element of a transition between the states $|\Psi_n \rangle$ and $|\Psi_k\rangle$ using the 
expression
\begin{eqnarray}
\langle D \rangle_{nk} &\equiv& \frac{\langle \Psi_n| D| \Psi_k\rangle}{\sqrt{\langle \Psi_n|\Psi_n\rangle\langle \Psi_k| \Psi_k\rangle}} \nonumber \\
&=& \frac{\langle\Phi_n|\tilde{D}_{nk}|\Phi_k\rangle}{\sqrt{\langle\Phi_n|\{1+\tilde{N}_n\}|\Phi_n\rangle 
\langle\Phi_k|\{1+\tilde{N}_k\}|\Phi_k\rangle}} , 
\end{eqnarray}
 where $\tilde{D}_{nk}=\{1+S_n^{\dagger} \} e^{T^{\dagger}} D e^T \{1+S_{k}\}$ and $\tilde{N}_{i=n,k}=\{1+S_i^{\dagger} \} e^{T^{\dagger}}
e^T \{1+S_{i}\}$. Calculation procedures of these expressions can be found elsewhere \cite{Mukherjee10,bijaya1,bijaya2,ref1,ref2,bijaya3}.

\section{Models for permittivity determination}~\label{model}

It is always desirable to have a logistic fit of the dielectric permittivity of a material media. For this purpose, a number of fitting models have
been proposed in the literature \cite{efelina2016,Mukherjee2015}. Drude developed a kinetic theory to account for the dielectric permittivity as 
well as its variation with frequency. In the Drude theory, the motion of a free electron in a material media can be described as a harmonic motion, 
where the electron oscillates under the influence of an electromagnetic wave. The oscillation leads to charge redistribution and create an additional
induced electric field that restores electrons to their equilibrium positions. This back and forth periodic motion of electrons can be described 
mathematically by oscillators. Within this harmonic oscillator model, the frequency-dependent permittivity~\cite{cai} can be presented as
\begin{equation}
\epsilon^{D}(\omega)=-\frac{\omega_{P}^{2}}{\omega^{2}+\iota\gamma_d \omega }, \label{drude}
\end{equation}
where $\omega_{P}$ is the plasma frequency relevant to the intraband transitions and can be written in terms of intrinsic carrier density $N$, reduced mass $m^*$ and permittivity of free space $\epsilon_0$  as
\begin{equation}
\omega_P = \frac{Ne^2}{\epsilon_0 m^*}\label{omegap}.
\end{equation}
Physically, the electromagnetic response of a material at $\omega_{P}$  changes from metallic  to dielectric. $\gamma_d$ in Eq.~\eqref{drude} is the
damping coefficient, which describes the damping force arising due to subsequent collisions of electrons and is expressed as
\begin{equation}
\gamma_d = \frac{e}{m^*\mu},\label{gammad}
\end{equation}
where $\mu$ is the carrier mobility and $e$ is the electron charge. In our calculations, its value is taken to be 0.041 m$^2$V$^{-1}$s$^{-1}$~\cite{Yan:15}. The Drude model 
describes contributions only from the free electrons to the permittivity, but it does not take into account the interband transitions of the bound 
electrons excited by the photons with higher energy. The contributions from these higher level interband electronic transitions to the dielectric 
permittivity can be expressed as a superposition of the Lorentz oscillators, given by 
\begin{equation}
\epsilon^{L}(\omega) = \sum_{j=1}^{5} \frac{f_{j} \omega_{P}^{2}}{\omega_{j}^{2}-\omega^{2}-\iota \gamma_{j} \omega},\label{dl}
\end{equation}  
where $j$ stands for the resonant nodes, $\omega_j$ corresponds to the resonance frequencies, $f_j$ refers to the weighting factor and $\gamma_j$ 
is the damping coefficient. It is worth noting that the Lorentz model reduces to the Drude Model for $j=0$, $\omega_0=0$, $f_{j}=1$ and $\gamma_0=\gamma_d$. 

In real materials, both free and bound electrons contribute to the dielectric permittivity. Therefore, the complete model contains both Drude 
component for intra band effect and Lorentz contribution for interband transitions. Accounting for them, the comprehensive DL model is 
represented as
\begin{equation}
\epsilon^{DL}(\omega) = \epsilon_{\infty}+\epsilon^{D}(\omega)+\epsilon^{L}(\omega),
\end{equation}
where $\epsilon_{\infty}$ is the permittivity at $\omega \rightarrow \infty$, denoting the constant offset value. We have used this model to fit 
the available experimental values given in Ref.~\cite{yu}, then infer values at other frequencies for their applications.

\begin{table}[t]
\caption{\label{pol1} Comparison of static polarizabilities (in a.u.) of the ground states of the Li, Na, K, Rb and Cs alkali atoms with their 
experimental values. Breakdown of different electron correlation effects for the determination of polarizabilities are also given explicitly.}
\begin{tabular}{lccccc}
 &  &  &  &  &  \tabularnewline
\hline 
\hline 
& & & \\
  & Li  & Na  & K  & Rb  & Cs  \tabularnewline
\hline 
& & & \\
Main  & 162.5 & 161.4  &  284.3 & 309.4 & 382.9 \tabularnewline
Core  & 0.2  &  0.9 &  5.5 & 9.1  & 15.8   \tabularnewline
Valence-core  & $\sim$0.0 & $\sim$0.0  & -0.1  & -0.3  & -0.5  \tabularnewline
Tail & $\sim$0.0 & 0.08 &  0.06 & 0.11&  0.15\tabularnewline
\hline
& & & \\
Total & 162.7 &   162.3& 289.8 &  318.5 & 398.4 \tabularnewline
Experimental& 164.2\cite{miffre}    &  162.7\cite{holmgren}  & 289.7\cite{greg}    & 319.8\cite{greg}   & 400.8\cite{greg} \tabularnewline 
\hline \hline
\end{tabular}
\end{table}

\begin{figure}
	\includegraphics[width=0.50\textwidth,height=0.34\textwidth]{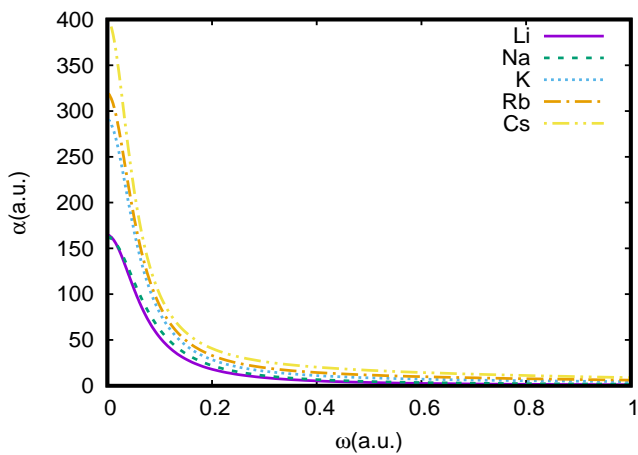}
	\caption{\label{fig:10} Plots showing dynamic polarizabilities (in a.u.) of the alkali Li, Na, K, Rb and Cs atoms in their ground state 
	as function of frequency (in a.u.).}
\end{figure}

\begin{table*}
\caption{\label{para} Fitting parameters for the dynamic polarizabilities ($\alpha_n(\iota\omega)$) of the considered alkali atoms. For unit 
conversion, one can use 1 a.u. of frequency $\omega$ = 27.21 eV and 1 a.u. of $\alpha_n(\iota\omega)$= 0.2488319 kHz 
(kV cm$^{-1}$ )$^{-2}$.}
\begin{center}
\begin{tabular}{|c|cc|cc|cc|cc|cc|}
\hline 
\hline
   & \multicolumn{10}{c|}{Atom} \\
   \cline{2-11}
   &\multicolumn{2}{c|}{Li}&\multicolumn{2}{c|}{Na}&\multicolumn{2}{c|}{K}&\multicolumn{2}{c|}{Rb}&\multicolumn{2}{c|}{Cs} \\
    \cline{2-11}
 Parameter & \multicolumn{10}{c|}{Frequency ($\omega$) in a.u.} \\
 \cline{2-11}
&  $0-1.85$ & $1.86-3000$& $0-2.24$ & $2.25-3000$ &$0-1.4$ & $1.5-3000$ &$0-1.05$ & $1.06-3000$ & $0-0.99$ & $1-3000$ \\
\hline
& & & & & & & && & \\
${\alpha}_0$ & 0.44277 &0.0005 &0.74571 &0.00005 & 4.28795 &0.0022 &7.26021  & 0.00813 &12.54507 & 0.01391\\ 
${\omega}_c$& -0.00032 & -0.91386  & -0.00013     & -1.82586  & -0.00028    & -0.64706  & -0.00041  & -0.70395 &-0.00062  & -0.6725\\ 
w &0.13766 & 1.75999 & 0.1555    & 2.83711   & 0.1201    &1.50212  & 0.11815 & 1.12351  & 0.10812 &1.2767   \\
A& 35.39279 & 11.55682& 39.47251& 28.46848&  53.87249 &60.89532  & 57.79795 & 110.35517 & 65.74528 & 137.66287\\            
\hline
\hline 
\end{tabular}
\end{center}
\end{table*}

\section{Results and Discussion}~\label{results} 

For realizing interactions between the multi-layered molybdenum disulfide with the alkali atoms, we require accurate values of dynamic 
polarizabilities of the alkali atoms. To validate the rigid correctness of these values, we first determine the static polarizabilities for 
the ground state of the considered alkali atoms and compare them with the available measurements. Our final calculated polarizability values 
along with the contributions from the core, core-valence and valence correlations are tabulated in Table \ref{pol1}. As can be seen, our calculated 
value of the ground state of Li is 162.7 a.u., which is in good agreement with the polarizability value of 164.2 a.u. measured by Miffre 
{\it et al.}\cite{miffre} using atom interferometry. Similarly, our estimated value for Na atom is 162.3 a.u. against its experimental result 
162.7 a.u. reported by Holmgren {\it et al.} \cite{holmgren}. The values obtained for K, Rb and Cs atoms from our calculations are 289.8 a.u., 
318.5 a.u. and 398.4 a.u., respectively. These values are also in good agreement with available measurements \cite{greg}. This demonstrates that 
the dynamic dipole polarizabilities of the investigated alkali atoms can be determined with sub-one percent accuracy for the intended study.

 We plot the dynamic polarizabilities obtained by us for the alkali atoms in Fig. \ref{fig:10}. To infer their values at a particular frequency, 
 we provide a fitting formula as
\begin{equation}
\alpha(\iota\omega)={\alpha}_0+\frac{2A}{\pi}\frac{\rm {w}}{4(\omega-{\omega}_c)^2+\rm{w}^2} ,
\end{equation}
where ${\alpha}_0$, $A$, ${\rm w}$ and ${\omega}_c$ are the fitting parameters. These parameters depend on the atom and range of frequency. We 
provide these fitting parameters in Table~\ref{para} for two different ranges of frequency to extrapolate the results.

\begin{figure}[htb!]
	\includegraphics[width=0.50\textwidth,height=0.34\textwidth]{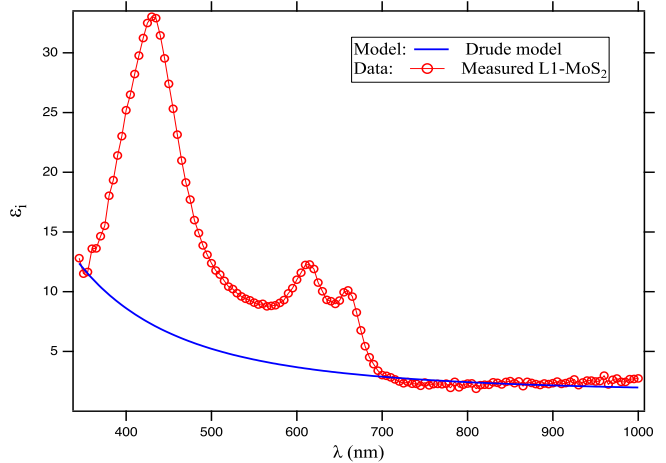}
	\caption{\label{fig:Drude} A comparative analysis of the imaginary part of the permittivity of the monolayer MoS$_{2}$ film estimated 
	using the Drude model as given by Eq.~(\ref{drudei}) (blue curve) and the measured spectra from Ref.~\cite{yu} (red circles). }
\end{figure}

\begin{figure}[htb!]
	\includegraphics[width=0.50\textwidth,height=0.34\textwidth]{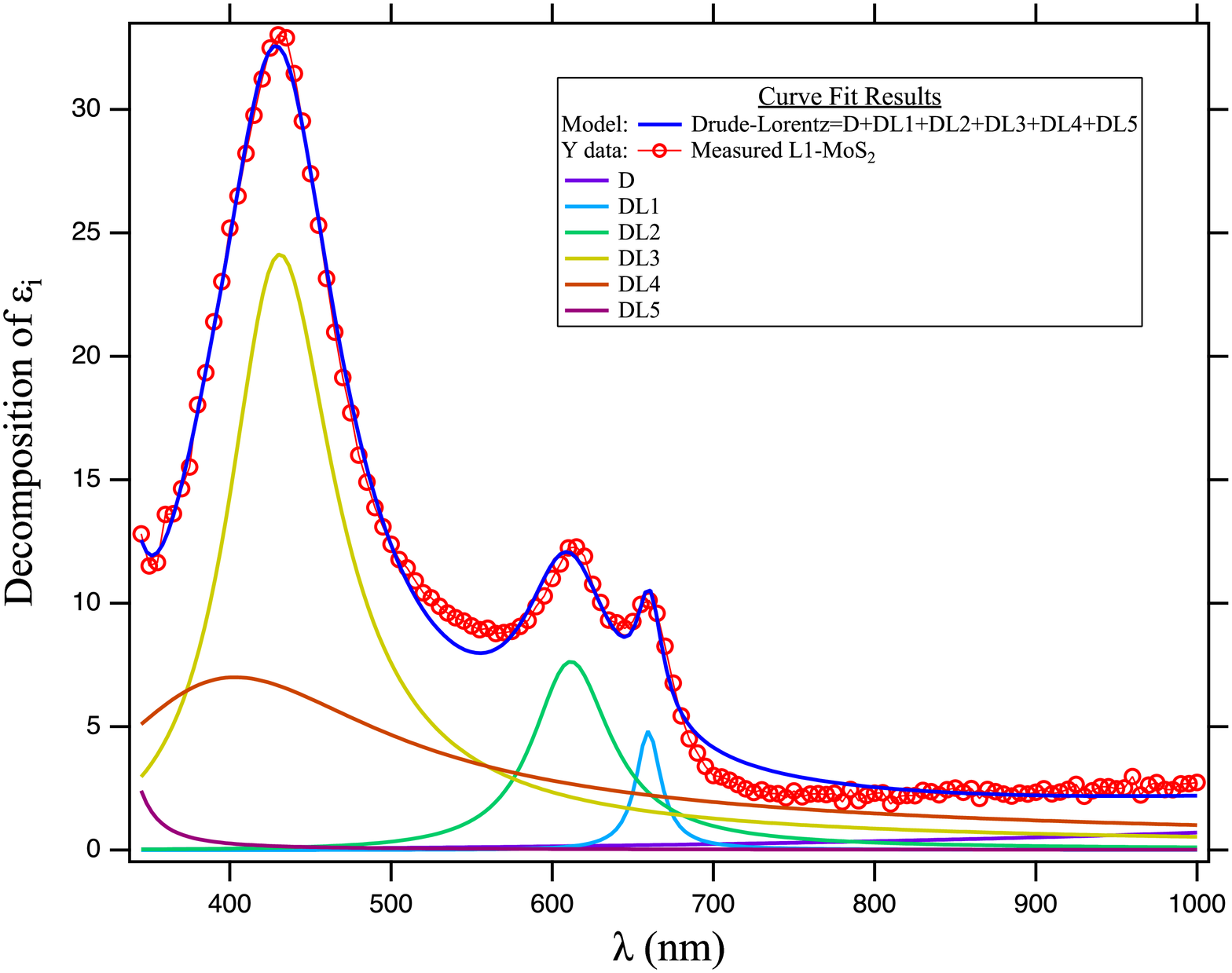}
	\caption{\label{fig:LD} Plots of the imaginary parts of the dynamic permittivity values of the monolayer MoS$_{2}$ film estimated using the 
	DL model given by Eq.~(\ref{dli}) (blue curve) and the measured spectra from Ref.~\cite{yu} (red circles) against wavelength
	(in nm). The $\epsilon_{i}$ values are decomposed into six components. The first component is named as `D' corresponding to the first term
	of Eq.~(\ref{dli}), whereas the other five components marked as `DL' corresponding to $j=$1,2,3,4 and 5 in the summation of Eq.~(\ref{dli}).}
\end{figure}

\begin{figure}[htb!]
	\includegraphics[width=0.50\textwidth,height=0.34\textwidth]{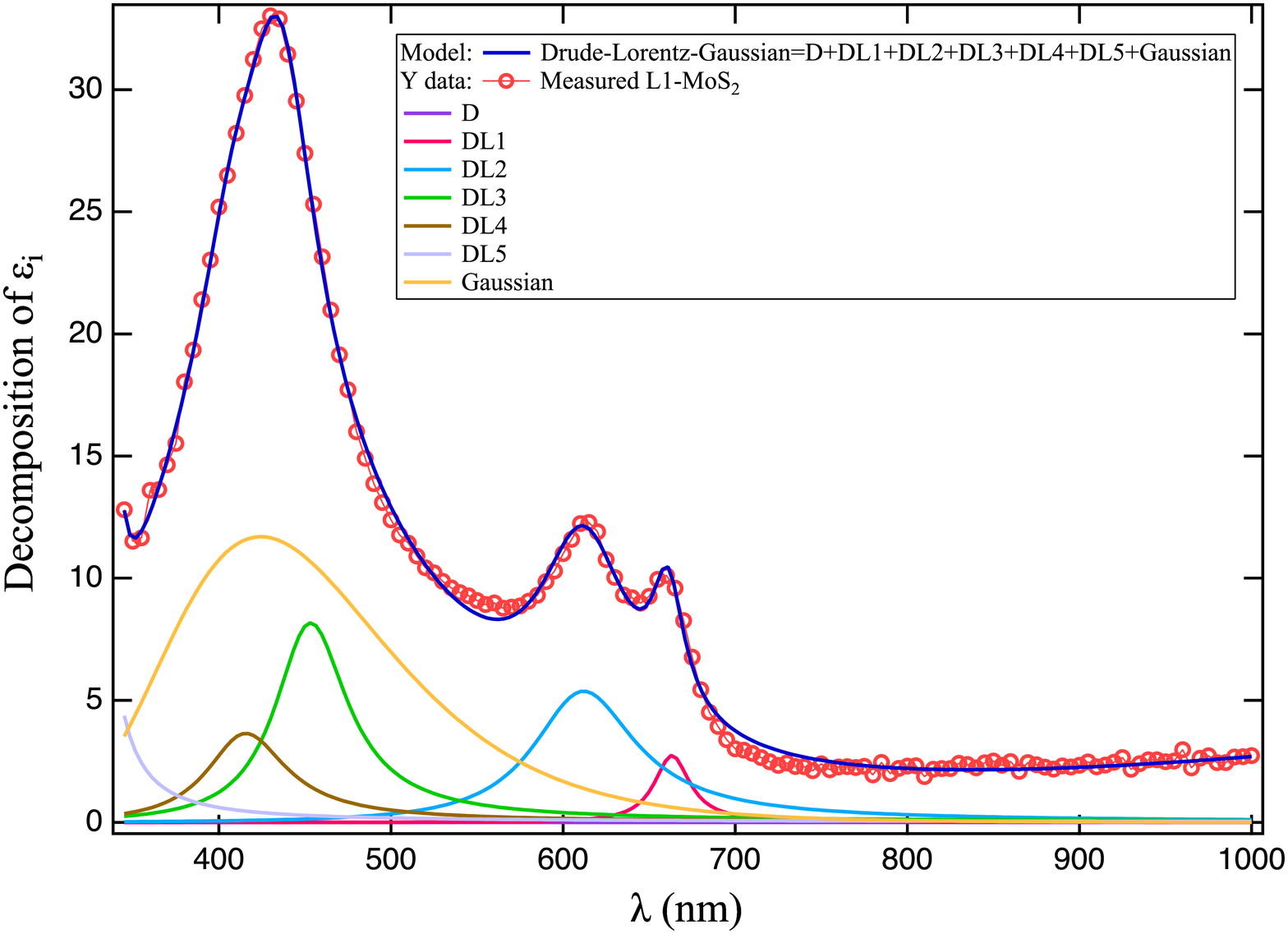}
	\caption{\label{fig:LDG} Plots of the imaginary parts of the permittivity of the monolayer MoS$_{2}$ film estimated by the 
	LDG model given by Eq.~\eqref{ldg} (blue curve) and from the measured spectra of Ref.~\cite{yu} (red circles) against wavelength
	(in nm). 
	Here, the $\epsilon_{i}$ is decomposed into seven components. The first component named as `D' corresponding to $j=$0 term of 
	Eq.~(\ref{ldg}), whereas the next five components marked as `DL' corresponding to $i=$1,2,3,4 and 5. The last term corresponds to 
	the Gaussian background added.}
\end{figure}

\begin{table*}

	\caption{\label{table:2a} Weight factor $f_{j}$ (dimensionless), damping coefficient $\gamma_{j}$(in eV), and $\omega_{j}$ resonance
		frequencies (in eV) for the Lorentz oscillators used in Eq.~\eqref{dli} for layers 1, 2 and 3. All the coefficients are normalized
		with $\hbar$  }
		\begin{tabular}{r| ccc| ccc| ccc|}
			\hline \hline 
			\multicolumn{1}{l|}{\multirow{2}{*}{$j$}} & \multicolumn{3}{c|}{Layer 1} & \multicolumn{3}{c|}{Layer 2}   & \multicolumn{3}{c}{Layer 3}\\ \cline{2-10} 
			\multicolumn{1}{l|}{}    & $f_{j}/\hbar[\times 10^{5}]$   &  $\gamma_{j}/\hbar[\times 10^{-1}]$  &  $\omega_{j}/\hbar$  & $f_{j}/\hbar[\times 10^{5}]$   &  $\gamma_{j}/\hbar[\times 10^{-1}]$ & $\omega_{j}/\hbar$ & $f_{j}/\hbar[\times 10^{5}]$   &  $\gamma_{j}/\hbar[\times 10^{-1}]$  &  $\omega_{j}/\hbar$  \\ 
			\hline 
			& & & & & & & & &  \\
			1 & $0.25\pm0.03$    &  $0.53\pm0.04$     &  $1.877\pm0.002$  &  $ {0.28 \pm 0.01}$     &  $ {0.64 \pm 0.08}$ & $1.869\pm0.002$ &  $ {0.21\pm0.05}$   &$ 0.68\pm0.03$       & $1.867 \pm 0.001$    \\ 
			2 & $1.85\pm0.22$    &   $2.22\pm0.13$    &  $2.034\pm0.002$  &  $1.51\pm 0.16$     &    $2.15\pm0.12$   &  $2.027\pm0.001$ &  $1.26\pm0.07$      & $ {2.14\pm0.09}$      &    $2.035 \pm 0.002$\\ 
			3 & $28.81\pm0.17$    &    $6.22\pm0.09$   &   $2.895\pm0.004$ &$   {20.39 \pm 0.31}$     &  $ {5.61\pm0.07}$     &  $ {2.858}\pm0.003$ &  $15.55\pm0.36$      & $ {5.48\pm0.14}$      & $2.812 \pm 0.004$   \\ 
			4 & $1.31\pm0.24$     &   $2.89\pm0.15$    & $3.19\pm0.03$   &  $ {5.48\pm0.83}$     &  $ {4.27\pm0.24}$     &  $ {3.16}\pm0.03$ &  $6.80\pm0.91$     &   $ {5.04\pm0.19}$    &  $3.15 \pm 0.04$  \\ 
			5 & $1.23\pm0.72$     &  $6.80\pm0.94$     &  $3.80\pm0.48$  & $3.6\pm1.2$      &  $3.30\pm0.66$     &  $3.76\pm0.41$ &  $3.46\pm0.98$    &   $ {3.9\pm1.3}$    &     $3.67 \pm 0.48$  \\ 
			\hline \hline
		\end{tabular}
\end{table*}

\begin{table*}
	
	\caption{\label{table:2b} Weight factor $f_{j}$ (dimensionless), damping coefficient $\gamma_{j}$(in eV), and $\omega_{j}$ resonance
		frequencies (in eV) for the Lorentz oscillators used in Eq.~\eqref{dli} for layers 4, 5 and 6. All the coefficients are normalized
		with $\hbar$  }
	\begin{tabular}{r| ccc| ccc| ccc|}
		\hline \hline 
		\multicolumn{1}{l|}{\multirow{2}{*}{$j$}} & \multicolumn{3}{c|}{Layer 4} & \multicolumn{3}{c|}{Layer 5}   & \multicolumn{3}{c|}{Layer 6}\\ \cline{2-10} 
		\multicolumn{1}{l|}{}    & $f_{j}/\hbar[\times 10^{5}]$   &  $\gamma_{j}/\hbar[\times 10^{-1}]$  &  $\omega_{j}/\hbar$  & $f_{j}/\hbar[\times 10^{5}]$   &  $\gamma_{j}/\hbar[\times 10^{-1}]$ & $\omega_{j}/\hbar$ & $f_{j}/\hbar[\times 10^{5}]$   &  $\gamma_{j}/\hbar[\times 10^{-1}]$  &  $\omega_{j}/\hbar$  \\ 
		\hline 
		& & & & & & & & &  \\
		1   &   $0.20\pm0.05$    &  $0.63\pm 0.06$     &    $1.866\pm0.001$ & $0.18\pm0.06$ & $0.52 \pm0.09$ & $1.859\pm0.001$ & $0.19\pm0.06$    &  $0.58\pm0.10$   &  $1.871\pm0.002$ \\ 
		2    &    $1.43\pm0.13$   &   $2.16\pm0.17$    & $2.038\pm0.001$ & $1.37\pm0.17$ & $2.11\pm0.21$ &   $2.028\pm0.002$ & $1.39\pm0.16$       &   $2.08\pm0.22$    &    $2.036\pm0.003$ \\ 
		3  &  $16.85\pm0.19$     &    $5.44\pm0.11$   &    $2.794\pm0.004$ & $15.03\pm0.42$ & $5.29\pm0.18$ & $2.761\pm0.002$ & $15.20\pm 0.34$      &    $5.03\pm0.14$   &     $2.757\pm0.005$ \\ 
		4      &   $7.63\pm0.82 $    & $4.95\pm0.18$      &  $3.13\pm0.06$& $8.12\pm0.85$ & $5.03\pm0.22$&  $3.10\pm0.04$ &   $8.68\pm0.77$    &   $4.78\pm0.19$    & $3.07\pm0.04$   \\ 
		5   & $5.6\pm2.2$       &     $4.6\pm1.4$  &   $3.62\pm0.43$ & $4.9\pm1.9$ & $4.6\pm1.4$ & $3.57\pm0.51$ &   $5.0\pm2.0$  &  $4.4\pm1.4$     &  $3.56\pm0.46$    \\ 
	
		\hline \hline
	\end{tabular}
\end{table*}

\begin{table*}
	
	\caption{\label{table:2c} Weight factor $f_{j}$ (dimensionless), damping coefficient $\gamma_{j}$(in eV), and $\omega_{j}$ resonance
		frequencies (in eV) for the Lorentz oscillators used in Eq.~\eqref{dli} for layers 7, 8 and 9. All the coefficients are normalized
		with $\hbar$  }
	\begin{tabular}{r| ccc| ccc| ccc|}
		\hline \hline 
		\multicolumn{1}{l|}{\multirow{2}{*}{$j$}} & \multicolumn{3}{c|}{Layer 7} & \multicolumn{3}{c}{Layer 8}   & \multicolumn{3}{c|}{Layer 9}\\ \cline{2-10} 
		\multicolumn{1}{l|}{}    & $f_{j}/\hbar[\times 10^{5}]$   &  $\gamma_{j}/\hbar[\times 10^{-1}]$  &  $\omega_{j}/\hbar$  & $f_{j}/\hbar[\times 10^{5}]$   &  $\gamma_{j}/\hbar[\times 10^{-1}]$ & $\omega_{j}/\hbar$ & $f_{j}/\hbar[\times 10^{5}]$   &  $\gamma_{j}/\hbar[\times 10^{-1}]$  &  $\omega_{j}/\hbar$  \\ 
		\hline 
		& & & & & & & & &  \\
		1 &  $0.23\pm0.05$     &  $0.58\pm0.05$     & $1.871\pm0.001$       &    $0.22\pm0.04$   &$0.59 \pm0.11$       & $1.867\pm0.002$   &  $0.28\pm0.07 $    &  $ {0.64\pm0.13}$     &    $1.873\pm0.004$     \\
		2 &  $1.52\pm0.11$     &    $2.05\pm0.21$   &  $2.038\pm0.003$     &  $1.62\pm0.11$      & $2.06\pm0.26$      &    $2.042\pm0.002$  &  $1.85\pm0.13$   &   $2.13\pm0.31$    & $2.039\pm0.004$     \\
		3 &  $17.05\pm0.29$     &  $5.10\pm0.21$     &  $2.749\pm0.004$     & $17.68\pm0.29 $      & $ {5.07}\pm0.31$      & $2.744\pm0.004$      &  $19.98\pm 0.31$     &    $5.20\pm0.39$   &    $2.721\pm0.005$    \\
		4 &  $8.43\pm0.93$      &   $4.73\pm0.31$    &   $3.08\pm0.05$    &  $8.80\pm1.07$     &   $4.80\pm0.25$    &  $3.06\pm0.06$   &  $7.98\pm0.97$    & $4.79\pm0.41$      &  $3.05\pm0.05$     \\ 
		5 &  $5.0\pm2.0$      &  $4.4\pm1.5$     &  $3.55\pm0.48$     &   $5.0\pm1.8$    &   $4.4\pm1.4$    &     $3.55\pm0.51$  &  $5.1\pm 1.8$       &     $4.4\pm1.4$  &   $3.56\pm0.72$  \\ 
		\hline \hline
	\end{tabular}
\end{table*}

\begin{table}
	\caption{\label{table:2d} Weight factor $f_{j}$ (dimensionless), damping coefficient $\gamma_{j}$(in eV), and $\omega_{j}$ resonance
		frequencies (in eV) for the Lorentz oscillators used in Eq.~\eqref{dli} for layers 9 and 10. All the coefficients are
		normalized with $\hbar$. }
		\begin{tabular}{r| ccc| }
			\hline \hline 
			\multicolumn{1}{l|}{\multirow{2}{*}{$j$}} & \multicolumn{3}{c|}{Layer 10}  \\ \cline{2-4} 
			\multicolumn{1}{l|}{}    & $f_{j}/\hbar[\times 10^{5}]$   &  $\gamma_{j}/\hbar[\times 10^{-1}]$  &  $\omega_{j}/\hbar$    \\ 
			\hline 
			& &  & \\
			1 & $0.25\pm0.06$ & $0.55\pm0.08 $ & $1.861\pm0.004$ \\
			2 & $1.68\pm0.12$ & $ {2.01\pm0.26}$ &   $2.037\pm0.005$ \\
			3& $18.17\pm0.34$ & $4.83\pm0.34$ & $2.72\pm0.005$ \\
			4 & $7.40 \pm1.10$ & $4.59\pm0.31$&  $3.05\pm0.05$ \\ 
			5  & $3.4\pm1.9$ & $4.1\pm1.6$ & $3.56\pm0.55$  \\ 
			\hline \hline
		\end{tabular}
\end{table}

\begin{table}
	\footnotesize
	\caption{\label{table:3} Values of the plasma frequency $\omega_{P}$, damping coefficient $\gamma_{d}$ and the calculated intrinsic carrier density $N$ for layer numbers 1 to 10. Here 
	$m_0$ is the mass of an electron. }
	\begin{tabular}{ c c c c c}
		\hline \hline
		\multicolumn{1}{c}{Layer} 	&\multicolumn{1}{c}{$\omega_{P}$}& \multicolumn{1}{c}{$\gamma_{d}$}               & \multicolumn{1}{c}{$m^{*}$} & \multicolumn{1}{c} {$N$}                               \\
		number	&\multicolumn{1}{c}{ (in meV)}    & \multicolumn{1}{c}{(in $ \times 10^{-2} $ eV)} &   \multicolumn{1}{c}{(in $m_0$)}     & \multicolumn{1}{c}{(in $\times 10^{15}$ cm$^{-3}$)}  \\ \hline
		& & & & \\
		1	& $27.59\pm0.02$  &$3.10\pm0.93$& $0.57\pm0.30$ &$1.20\pm 0.63$\\ 
		2	& $26.68\pm0.02$ &$3.18\pm0.81$ &$0.56\pm0.25$  &$1.13\pm  0.52$\\
		3	&  $25.59\pm0.01$ &$3.20\pm0.78$& $0.55\pm0.24$ &$1.08\pm0.48$\\
		4	&  $25.16\pm0.02$ &$3.26\pm0.82$& $0.54\pm0.25$ &$1.04\pm 0.49$\\
		5	&  $25.06\pm0.03$&$3.31\pm0.86$ & $0.54\pm0.26$ &$1.02\pm0.49$\\
		6	& $24.81\pm0.02$  &$3.42\pm0.77$&$0.52\pm 0.23$ &$0.98\pm 0.43$\\
		7	&  $25.31\pm0.04$&$3.39\pm1.05$ & $0.52\pm0.31$ &$1.01 \pm 0.60$\\
		8	&  $25.63\pm0.02$&$3.33\pm0.81$& $0.53\pm 0.24$ &$1.04\pm0.48$\\
		9	&  $25.96\pm0.02$ &$3.28\pm0.94$&$0.54\pm0.29 $ &$1.07\pm0.56$\\
		10      &  $26.34\pm0.04$&$3.20\pm0.77$ &$0.55\pm0.24$  &$1.11\pm0.49$\\
		\hline \hline
	\end{tabular}
\end{table}

We use the previously discussed models to fit the dynamic values of permittivity available in literature~\cite{yu} and recommend the best fitted 
permittivity model for MoS$_2$ layers. We consider only the imaginary part of the permittivity as the real part can be estimated using the 
Kramer-Kronig relation, given by Eq.~{\eqref{eq:9}}. The formulae for the imaginary part of $\epsilon$ following Eqs.~(\ref{drude}) and
(\ref{dl}), are given by
\begin{equation}
\epsilon^{D}_i(\omega) = \frac{\gamma_d \omega_{P}^{2}}{\omega(\omega^{2}+\gamma_d^{2})}\label{drudei}
\end{equation}
and
\begin{equation}
\epsilon^{DL}_i(\omega) =  \frac{\alpha \gamma_{d} \omega_{P}^{2} }{\omega (\omega^{2}+\gamma_{d}^{2})} + \sum_{j=1}^{5} \frac{f_{j} \gamma_{j} \omega \omega_{P}^{2}}{\gamma_{j}^{2} \omega^{2} + (\omega^{2}-\omega_{j}^{2})^{2}} \label{dli}
\end{equation}
in the Drude and DL models, respectively. The $\epsilon^{D}_i(\omega)$ values of monolayer MoS$_2$  using Eq.~(\ref{drudei}) have been
graphically presented in Fig. \ref{fig:Drude}. As can be verified from the figure that the Drude model gives accurate permittivity values only
in the infrared region and the experimental data disagrees with predictions from Drude model in the visible wavelength (shorter than 700 nm). 
In this region various interband transitions start contributing. Therefore, it is expected that the DL model will provide a better fit to 
the measured values. The $\epsilon^{DL}_i(\omega)$ values using Eq.~(\ref{dli}) along with the experimental permittivity values are shown in 
Fig.~\ref{fig:LD}. It can be noticed from this figure that the measured data from Ref.~\cite{yu} is consistent with the results estimated by 
the DL model. This suggests that the values estimated using the DL model can be assumed to be reliable for further analysis.
The authors in Ref.~\cite{Mukherjee2015} have used a hybrid Lorentz-Drude-Gaussian (LDG) model in their study to fit the permittivity data 
for monolayer of MoS$_2$, which is given by
\begin{eqnarray}
\epsilon^{LDG}_i(\omega) &=& \epsilon_{\infty}+\sum_{j=0}^{5}\frac{f_j\gamma_j\omega_p^2\omega}{(\omega_j^2-\omega^2)^2+\omega^2\gamma_j^2}\nonumber \\&+& \eta\exp \left( -\frac{(\hbar \omega - \beta)^{2}}{2\sigma^{2}}\right), \label{ldg}
\end{eqnarray}
with $\beta$ as mean, $\sigma$ as variance and $\eta$ as the maximum amplitude of the Gaussian function. In this case, the term with $j=0$ and $\omega_0=0$ 
carries a weight factor $f_0\ne 1$. As seen 
in Fig.\ref{fig:LDG}, by adding a Gaussian background of the above kind with our DL model does not bring much change to our fitted values. Since 
there is no physical interpretation of the Gaussian component and it is added only as a background to match the estimated values with the 
experimental results, this justifies our above assertion that the DL model is able to predict permittivity values accurately.

In Tables \ref{table:2a}, \ref{table:2b}, \ref{table:2c}, \ref{table:2d}, and \ref{table:3} we present the fitting values of $\gamma_d$, 
$\omega_{P} $, $\gamma_j$, $\omega_j$ and $f_j$ from the DL model for various layers of MoS$_2$ along with the uncertainties in them. 
We use a non-linear least-squares minimization technique to extrapolate the permittivity values at different frequencies and they are decomposed into six components. The first component is named as `D' corresponding to the first term of Eq.~(\ref{dli}), whereas the other five components marked as `DL$_j$' corresponding to $j=$1,2,3,4 and 5 in the summation of Eq.~(\ref{dli}). We have provided a code written using the python programming language in the Supplementary Material that is used for carrying out this fitting. As can be seen from Table \ref{table:2a}, 
our $\omega_j/\hbar$ values for monolayer agree well with those predicted in literature~\cite{frindt,eda,mono1,mono2,mono3,Mukherjee2015}. 
We also note from these tables that although the uncertainties are quite small for the DL1, DL2, DL3 and DL4 components, they are quite significant 
for the DL5 and D components. For the DL5 component, only a few data points are available for fitting which lead to significant uncertainties in the 
fitting parameters corresponding to this component. Also, the inferred $\omega_p$ value of 27.59 meV for monolayer of MoS$_2$, given in Table \ref{table:3},
matches very well with the measured plasma frequency of 28.3 meV ~\cite{Shen_2013}.  We use the fitting value of $\gamma_d$ in Eq.~(\ref{gammad}) to 
estimate the effective mass $m^*$ for various number of MoS$_2$ layers. From these calculated $m^*$ values, we further evaluate intrinsic carrier density 
$N$ using the fitting values of $\omega_{P}$ in Eq.~(\ref{omegap}). From Table \ref{table:3}, we also note that $N$ is maximum for a monolayer of MoS$_2$, 
thereafter, it starts decreasing up to layer number 6 and starts increasing again as the number of layers are increased up to 10. 

\begin{figure}
	\includegraphics[scale=0.7]{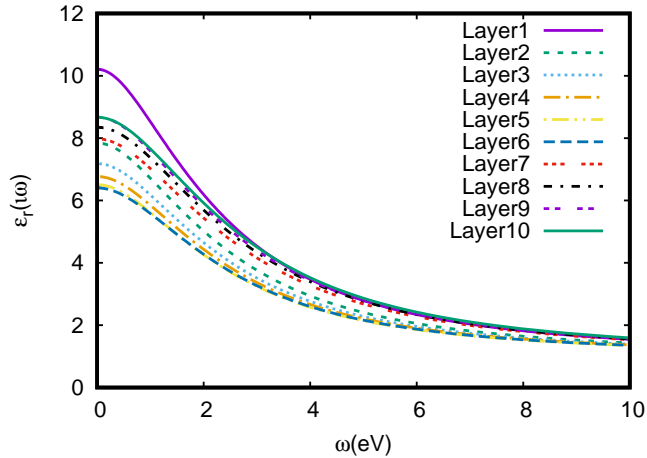}
	\caption{\label{3}Plots of real parts of the dielectric permittivity at imaginary frequency as function of frequencies (in eV) for different
	number of MoS$_{2}$ layers.}
\end{figure}

\begin{table}[t]
	\caption{\label{tab:I}Calculated C$_{3}$ coefficients (in a.u.) for interaction between different layers of MoS$_2$ with the alkali-metal atoms.}
	\begin{tabular}{clllll}
		&  &  &  &  &  \tabularnewline
		\hline 
		\hline 
		\multicolumn{1}{m{1cm}}{Layer}  & \multicolumn{1}{m{1cm}}{Li}  & \multicolumn{1}{m{1cm}}{Na}  & \multicolumn{1}{m{1cm}}{K}  & \multicolumn{1}{m{1cm}}{Rb}  & \multicolumn{1}{m{1cm}}{Cs}  \tabularnewline
		\hline 
1 & 0.879	&0.960	&1.455	&1.613	&1.932\tabularnewline
2 & 0.810	&0.883	&1.340	&1.484	&1.776\tabularnewline
3 & 0.784	&0.853	&1.296	&1.434	&1.716\tabularnewline
4 & 0.766	&0.833	&1.266	&1.401	&1.677\tabularnewline
5 & 0.752	&0.817	&1.243	&1.374	&1.645\tabularnewline
6 & 0.752	&0.818	&1.243	&1.375	&1.645\tabularnewline
7 & 0.839	&0.917	&1.388	&1.540	&1.845\tabularnewline
8 & 0.853	&0.933	&1.413	&1.567	&1.878\tabularnewline
9 & 0.860	&0.940	&1.424	&1.580	&1.893\tabularnewline
10& 0.866	&0.948	&1.435	&1.592	&1.908\tabularnewline
		\hline 
	\end{tabular}
\end{table}

\begin{figure}
	\includegraphics[scale=0.7]{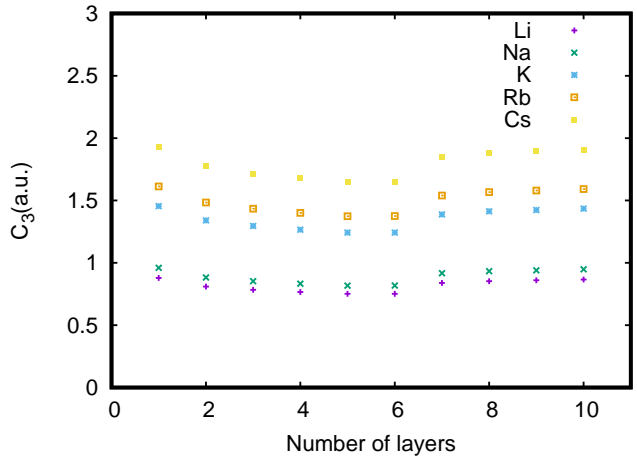}\caption{\label{fig:5}Plots showing the C$_3$ coefficients for Li, Na,
		K, Rb and Cs atoms with varying layer numbers of MoS$_2$.}
\end{figure}
 
Next, we find $\epsilon_r(\iota\omega)$ values extracted by substituting $\epsilon_i(\omega)$ values in  Eq.~\eqref{eq:9} for different
layers of MoS$_2$, which are plotted against frequency in Fig. \ref{3}. The behaviour of $\epsilon_r(\omega)$ as a function of layer number is seen 
to be in accordance with the observation by Yu et al.~\cite{yu}. In their work, the authors demonstrate that excitonic effects play a dominant 
role in the dielectric function of  5-7 layered MoS$_2$. Therefore, the dielectric function decreases with the layer number up to 6 but turns to 
increase further with the increase in layer number. 

\begin{figure}
	\includegraphics[scale=0.7]{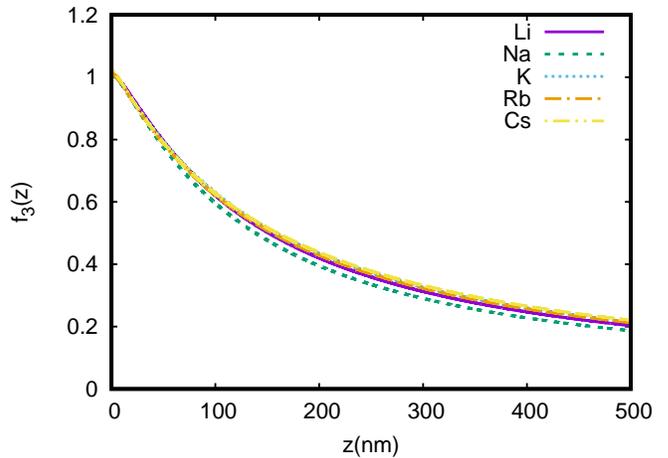}\caption{\label{fig:4}The retardation coefficients ($f_{3}(z)$) for Li, Na,
		K, Rb and Cs atom as functions of the atom-wall separation distance $z$.}
\end{figure}

The C\textsubscript{3} coefficients for the interactions between Li, Na, K, Rb and Cs atoms and the MoS\textsubscript{2} layers evaluated using 
$\epsilon_r(\iota\omega)$ values are listed in Table \ref{tab:I}. A comparison of the C$_3$ coefficient as a function of layer number reveals 
that the interaction is maximum between atoms and monolayer of MoS$_2$.  The interaction decreases with an increase in the number of MoS$_2$ layers 
up to the sixth layer, then it starts increasing again. The trend is found to be common for all the considered atoms. It is also quite evident 
that the trend followed by the C$_3$ coefficients with increasing number of layers is similar to that predicted for the intrinsic carrier 
density $N$. This observation is explained using the fact that the strength of the vdW force depends on the electric polarizability 
of the interacting atom. The tendency of the MoS$_2$ layer to polarize the incoming atom increases with the increase in the number of electrons 
per unit volume. As a result, the values of C$_3$ see an upsurge with an escalation in $N$.

 A graphical representation for the C$_3$ coefficients for the considered alkali atoms with varying layer number is shown in Fig.~\ref{fig:5}.  
Our results in this figure support the finding that for the same layer number, the C$_3$ coefficients increase with increase in the atomic 
number. We also notice that the ratio of C$_3$ coefficients among various atoms vary slowly with the number of layers. For instance, the ratio
of the C$_3$ coefficient for the interaction of any layer of MoS$_2$ with Rb and Li atoms is 1.83 irrespective of the number of layers. This
knowledge of variation pattern of C$_3$ coefficients with number of layers with different alkali atoms will pave way to design sensors for 
detecting the alkali atoms by the MoS$_2$ layers. To give an estimate of these interactions at an intermediate distance, we next calculate the
retardation function $f_{3}(z)$ as a function of distance $z$ for various number of layers in MoS$_2$ based TMDs. We have shown comparison of 
the $f_{3}(z)$ values between an atom and the MoS$_2$ monolayer in Fig. \ref{fig:4}. It is clear from this figure that the retardation function 
decreases with increase in $z$. Also, we notice from the above figure that the retardation function is similar for all the considered atoms, and 
it is not affected much with the atomic size. 

\section{Conclusion}

To summarize, we have investigated the C$_3$ coefficients for the interactions between the alkali atoms with the MoS$_2$ layers. We performed high 
accuracy calculations of dynamic dipole polarizability of the considered alkali atoms and determined the dynamic dielectric permittivities for
different layers of MoS$_2$ over a wide range of imaginary frequency. We have proposed a readily usable logistic fit for the dielectric permittivity
for various layers of MoS$_2$ ranging from 1  to 10. We have also shown dependency of the intrinsic carrier density $N$ and the coefficients with 
increasing layer numbers of the MoS$_2$ surface. Variation of C$_3$ as well as $N$ with the number of layers shows decrease in values up to 6 number 
of layers. This finding could be useful for the formation of highly sensitive  and reproducible sensing probes for detection of alkali atoms using 
1-6 layered MoS$_2$ based transition metal dichalcogenides. Our study reveals that the ratios of the C$_3$ coefficients among various atoms do not 
change as the layer number is changed.

\begin{acknowledgments}
The work of B.A. is supported by the DST-SERB Grant No. EMR/2016/001228. 
\end{acknowledgments}



\end{document}